\documentclass[twoside]{dis08}
\usepackage[latin1]{inputenc}
\usepackage[dvips]{graphicx,epsfig,color}
\usepackage{wrapfig,rotating}
\usepackage{amssymb,amsmath,array}

\begin{document}
\title{Recent Progress in Global PDF Analysis}

%***********************************************************************
% AUTHORS INFORMATION AREA
%***********************************************************************
\author{\underline{G.~Watt}$^1$, A.~D.~Martin$^2$, W.~J.~Stirling$^2$ and R.~S.~Thorne$^1$
%
% DO NOT MODIFY THE FOLLOWING '\vspace' ARGUMENT
\vspace{.3cm}\\
1- Department of Physics \& Astronomy, University College London, WC1E 6BT, UK
\vspace{.1cm}\\
2- Institute for Particle Physics Phenomenology, University of Durham, DH1 3LE, UK
}
%***********************************************************************
% END OF AUTHORS INFORMATION AREA
%***********************************************************************

\maketitle

\begin{abstract}
  We discuss selected topics in the forthcoming MSTW 2008 determination of parton distributions by global analysis.  The tolerance parameter controlling the uncertainties on the parton distributions is now determined by a new dynamic procedure for each eigenvector of the covariance matrix.  New data sets fitted include Tevatron Run II data on inclusive jet production, the lepton charge asymmetry from $W$ decays and the $Z$ rapidity distribution.  Predictions are given for the total $W$ and $Z$ cross sections at the Tevatron and LHC.
\end{abstract}

This talk~\cite{url} covers recent developments in the imminent MSTW 2008 global analysis, which is a major update to the currently available MRST 2001 LO~\cite{Martin:2002dr}, MRST 2004 NLO~\cite{Martin:2004ir} and MRST 2006 NNLO~\cite{Martin:2007bv} parton distribution functions (PDFs).

Much effort has been devoted in recent years towards providing PDFs with uncertainties obtained by propagating the experimental errors on the data sets included in the global fit.  The most widely used procedure is the \emph{Hessian} method pioneered by the CTEQ group.  The Hessian matrix $H$ is the matrix of second derivatives of the $\chi^2_{\rm global}$ with respect to the PDF parameters.  The covariance matrix $C\equiv H^{-1}$ is diagonalised to give a set of eigenvalues $\lambda_k$ and orthonormal eigenvectors $v_{ik}$.  The CTEQ and MRST fitting groups have produced eigenvector PDF sets $S_k^{\pm}$ with parameters $\{a_i\}$ shifted from the values at the global minimum: $a_i(S_k^\pm) = a_i^0 \pm t\,\sqrt{\lambda_k}v_{ik}$, with $t$ adjusted to give the desired \emph{tolerance} $T=(\Delta\chi^2_{\rm global})^{1/2}$.  Then PDF users can calculate uncertainties on a quantity $F(\{a_i\})$ with $(\Delta F)^2 = (1/4)\sum_k \left[F(S_k^+)-F(S_k^-)\right]^2$, or using a slightly more complicated formula to give asymmetric uncertainties.

Ideally, with the standard \emph{parameter-fitting} criterion, we would expect the errors to be given by the choice of tolerance $T=1$ for the $68\%$ (1-$\sigma$) confidence level (C.L.) limit, or $T=\sqrt{2.71}$ for the $90\%$ C.L.~limit.  This is appropriate if fitting consistent data sets with ideal Gaussian errors to a well-defined theory.  However, in practice, there are minor inconsistencies between the independent fitted data sets, and unknown experimental and theoretical uncertainties, so this criterion is not appropriate for global PDF analyses.  Instead, the much weaker \emph{hypothesis-testing} criterion has been used by the CTEQ and MRST groups, such that the eigenvector PDF sets are treated as alternative hypotheses.  The tolerance is then determined from the condition that each data set should be described within its 90\% C.L.~limit, giving approximate values of $T=10$ (CTEQ) or $T=\sqrt{50}$ (MRST), instead of the canonical parameter-fitting value of $T=\sqrt{2.71}$ for the 90\% C.L.~limit.

The choice of tolerance has been studied more quantitatively in the new MSTW 2008 analysis.  The 90\% C.L. region for each data set $n$ (with $N_n$ data points) is defined as $\chi_n^2<(\chi_{n,0}^2/\xi_{50})\,\xi_{90}$, where $\chi_{n,0}^2$ is the value of the goodness-of-fit measure for data set $n$ at the \emph{global} minimum, $\xi_{90}$ is the 90th percentile of the $\chi^2$-distribution with $N_n$ degrees of freedom, and $\xi_{50}\simeq N_n$ is the most probable value of the $\chi^2$-distribution with $N_n$ degrees of freedom.  The tolerance is then determined separately for each eigenvector direction to ensure that all data sets are described within their 90\% C.L.~limits.

\begin{figure}[t]
  \centering
  \includegraphics[width=\textwidth]{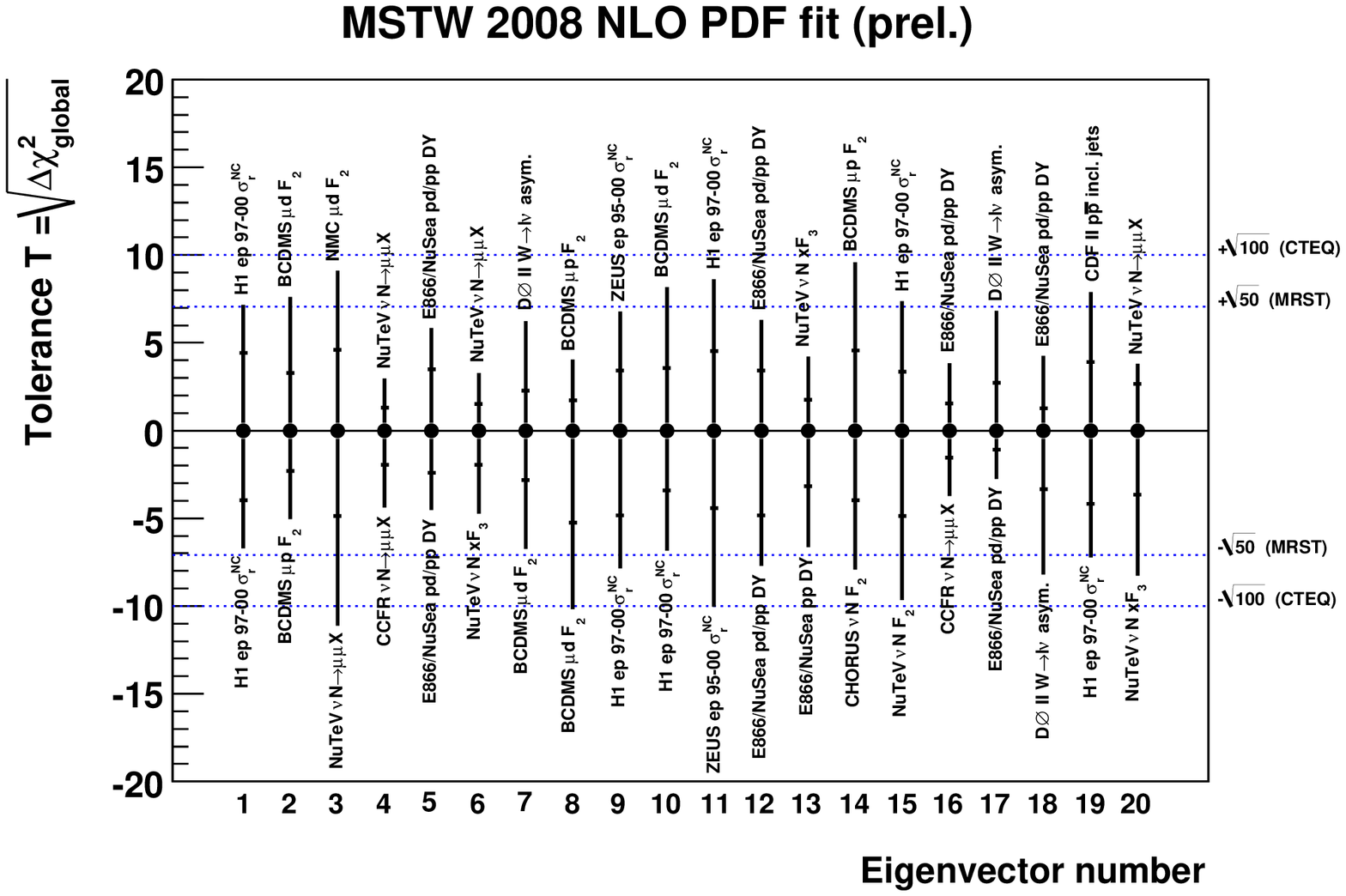}
  \vspace*{-8mm}
  \caption{The value of the tolerance parameter, $T=(\Delta\chi^2_{\rm global})^{1/2}$, determined dynamically for each eigenvector direction from the condition that each data set must be described within its $90\%$ C.L.~limit (outer error bars) or $68\%$ C.L.~limit (inner error bars).  The labels give the name of the data set which sets the 90\% C.L.~tolerance for each eigenvector direction.}
  \label{fig:tolerance}
\end{figure}
The $\overline{\textrm{MS}}$ PDF parameterisation at the input scale $Q_0^2 = 1$ GeV$^2$ is taken as:
\begin{align}
  xu_v &= A_u\,x^{\mbox{\large $\boldsymbol{\color{red} \eta_1}$}} (1-x)^{\mbox{\large $\boldsymbol{\color{red} \eta_2}$}} (1 + \mbox{\Large $\boldsymbol{\color{red} \epsilon_u}$}\,\sqrt{x} + \gamma_u\,x), \\
  xd_v &= A_d\,x^{\mbox{\large $\boldsymbol{\color{red} \eta_3}$}} (1-x)^{\mbox{\large $\boldsymbol{\color{red} \eta_4}$}} (1 + \mbox{\Large $\boldsymbol{\color{red} \epsilon_d}$}\,\sqrt{x} + \gamma_d\,x), \\
  xS & \equiv 2x\bar{u}+2x\bar{d}+xs+x\bar{s} = \mbox{\Large $\boldsymbol{\color{red} A_S}$}\,x^{\delta_S} (1-x)^{\mbox{\large $\boldsymbol{\color{red} \eta_S}$}} (1 + \mbox{\Large $\boldsymbol{\color{red} \epsilon_S}$}\,\sqrt{x} + \gamma_S\,x), \\
  x\Delta & \equiv x\bar{d} - x \bar{u} = \mbox{\Large $\boldsymbol{\color{red} A_\Delta}$}\,x^{\mbox{\Large $\boldsymbol{\color{red} \eta_\Delta}$}} (1-x)^{\eta_S+2} (1 + \mbox{\Large $\boldsymbol{\color{red} \gamma_\Delta}$}\,x + \delta_\Delta\,x^2), \\
  xg &= A_g\,x^{\mbox{\large $\boldsymbol{\color{red} \delta_g}$}} (1-x)^{\mbox{\large $\boldsymbol{\color{red} \eta_g}$}} (1 + \epsilon_g\,\sqrt{x} + \gamma_g\,x) + A_{g^\prime}\,x^{\mbox{\large $\boldsymbol{\color{red} \delta_{g^\prime}}$}} (1-x)^{\mbox{\large $\boldsymbol{\color{red} \eta_{g^\prime}}$}}, \label{eq:gluon} \\
  xs + x\bar{s} &= \mbox{\Large $\boldsymbol{\color{red} A_{+}}$}\,x^{\delta_S}\,(1-x)^{\mbox{\large $\boldsymbol{\color{red} \eta_{+}}$}} (1 + \epsilon_S\,\sqrt{x} + \gamma_S\,x), \label{eq:splus} \\
 xs - x\bar{s} &= \mbox{\Large $\boldsymbol{\color{red} A_{-}}$}\,x^{\delta_{-}} (1-x)^{\mbox{\large $\boldsymbol{\color{red} \eta_{-}}$}} (1-x/x_0). \label{eq:sminus}
\end{align}
The parameters $A_u$, $A_d$, $A_g$ and $x_0$ are fixed by enforcing number- and momentum-sum rule constraints, while the other parameters are allowed to go free.  The 20 highlighted parameters are those allowed to go free when producing the eigenvector PDF sets, where the other parameters are fixed.  This is to be compared with the 15 parameters for the MRST eigenvector PDF sets.  The 5 new parameters include an additional 4 parameters associated with the new strangeness degrees of freedom, \eqref{eq:splus} and \eqref{eq:sminus}, warranted by the inclusion of NuTeV/CCFR dimuon cross sections, and an additional free parameter in the input gluon distribution \eqref{eq:gluon} which allows more flexibility.  The tolerance for each of the 20 eigenvector directions determined dynamically using the above procedure is shown in Fig.~\ref{fig:tolerance}.  The tolerance is generally close to the MRST value of $\sqrt{50}$; however, it is significantly smaller for some eigenvectors, such as those associated with the strange quark parameters, which are mainly determined only from the NuTeV/CCFR dimuon cross sections.

\begin{figure}
  \centering
  (a)\hspace{0.45\textwidth}(b)\hspace*{0.45\textwidth}\ \\
  \includegraphics[width=0.5\textwidth]{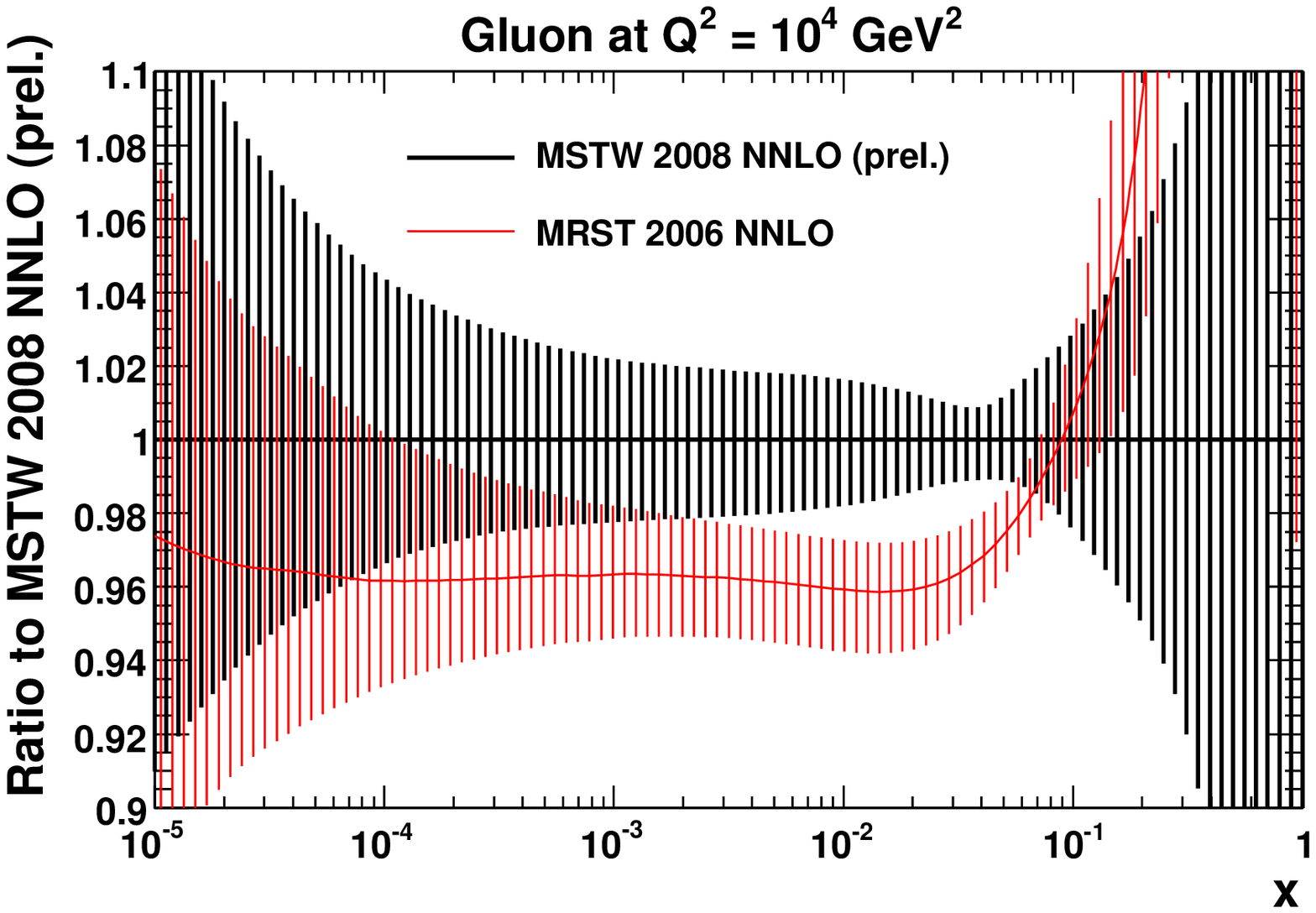}%
  \includegraphics[width=0.5\textwidth]{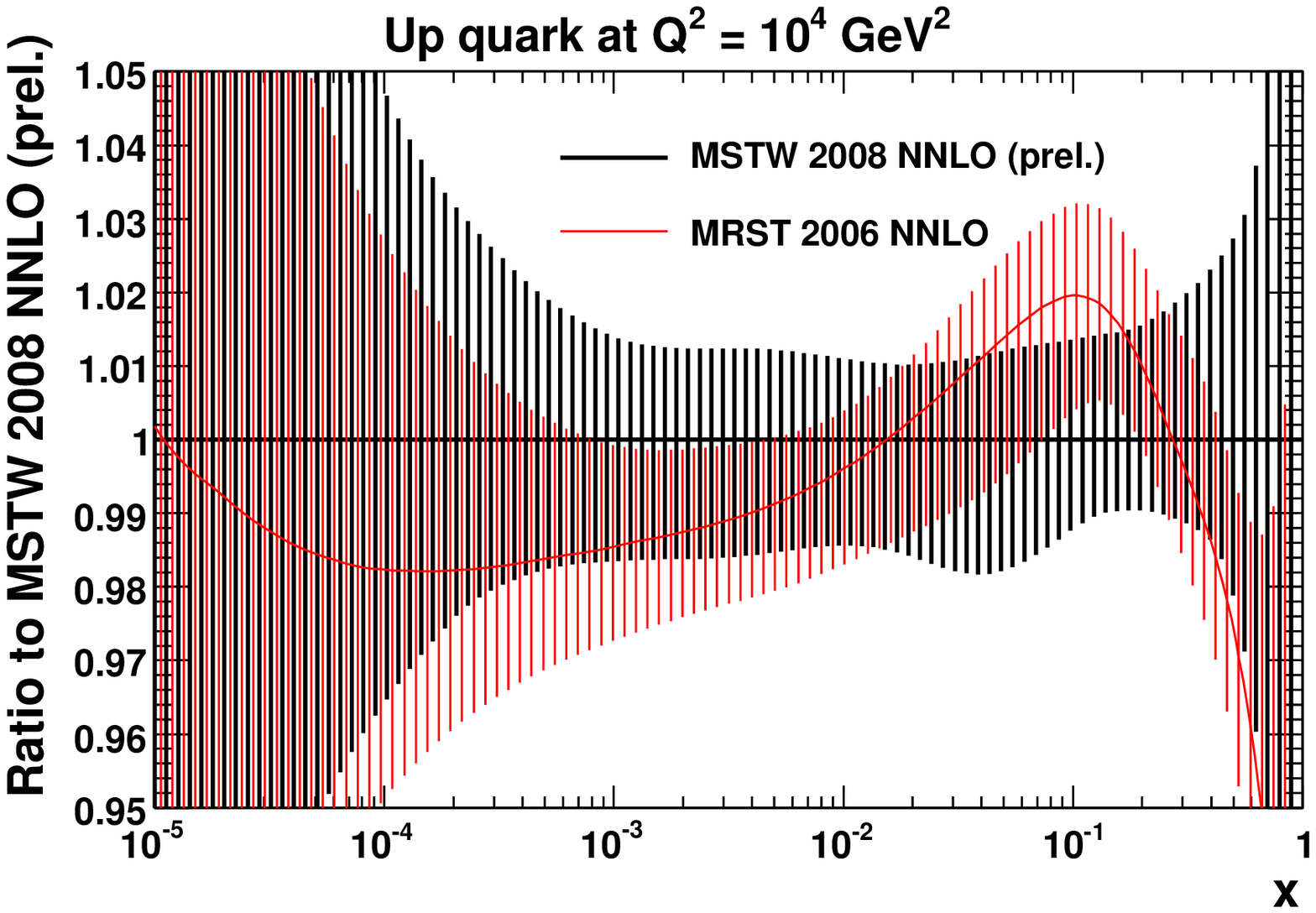}\\
  (c)\hspace{0.45\textwidth}(d)\hspace*{0.45\textwidth}\ \\
  \includegraphics[width=0.5\textwidth]{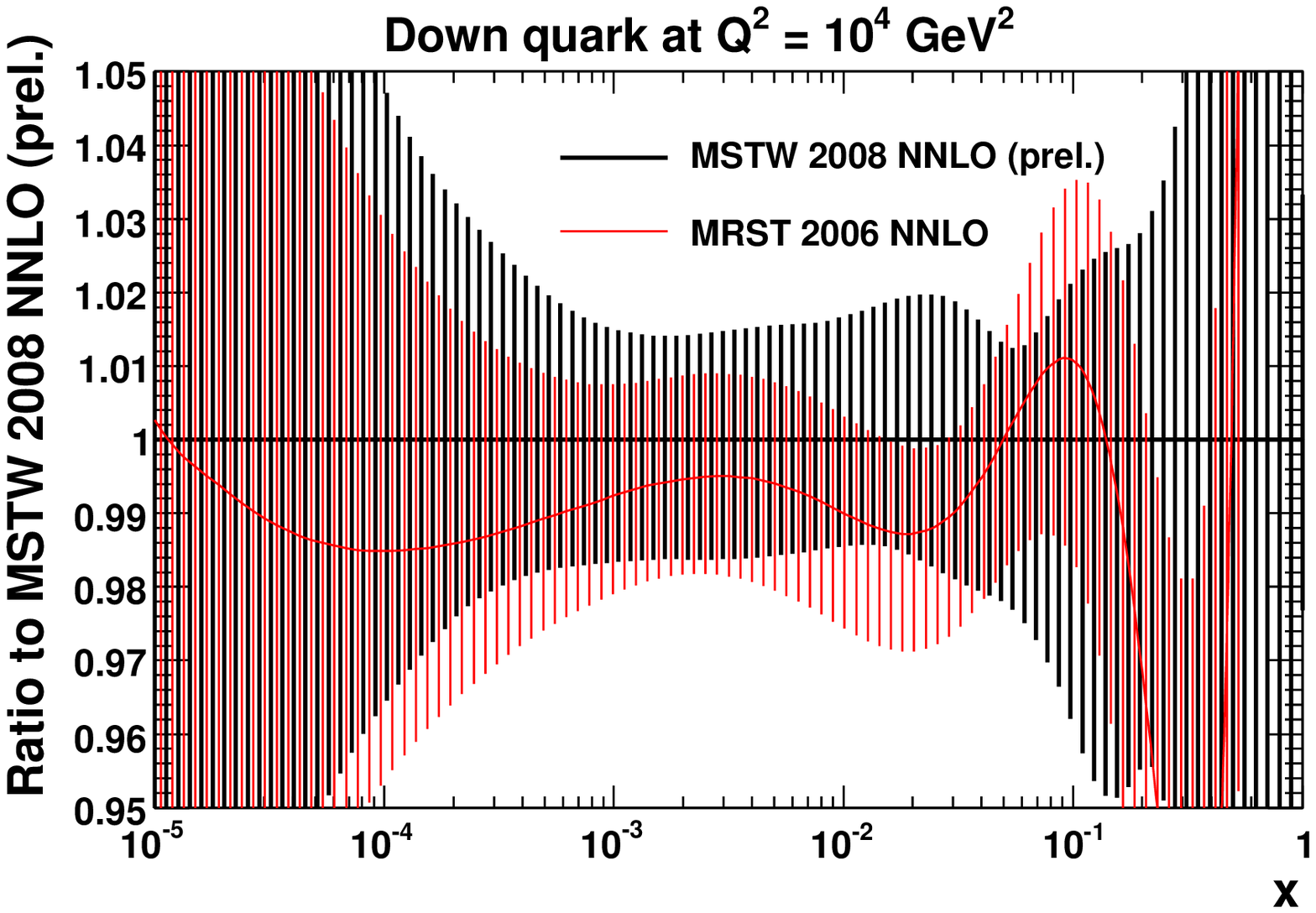}%
  \includegraphics[width=0.5\textwidth]{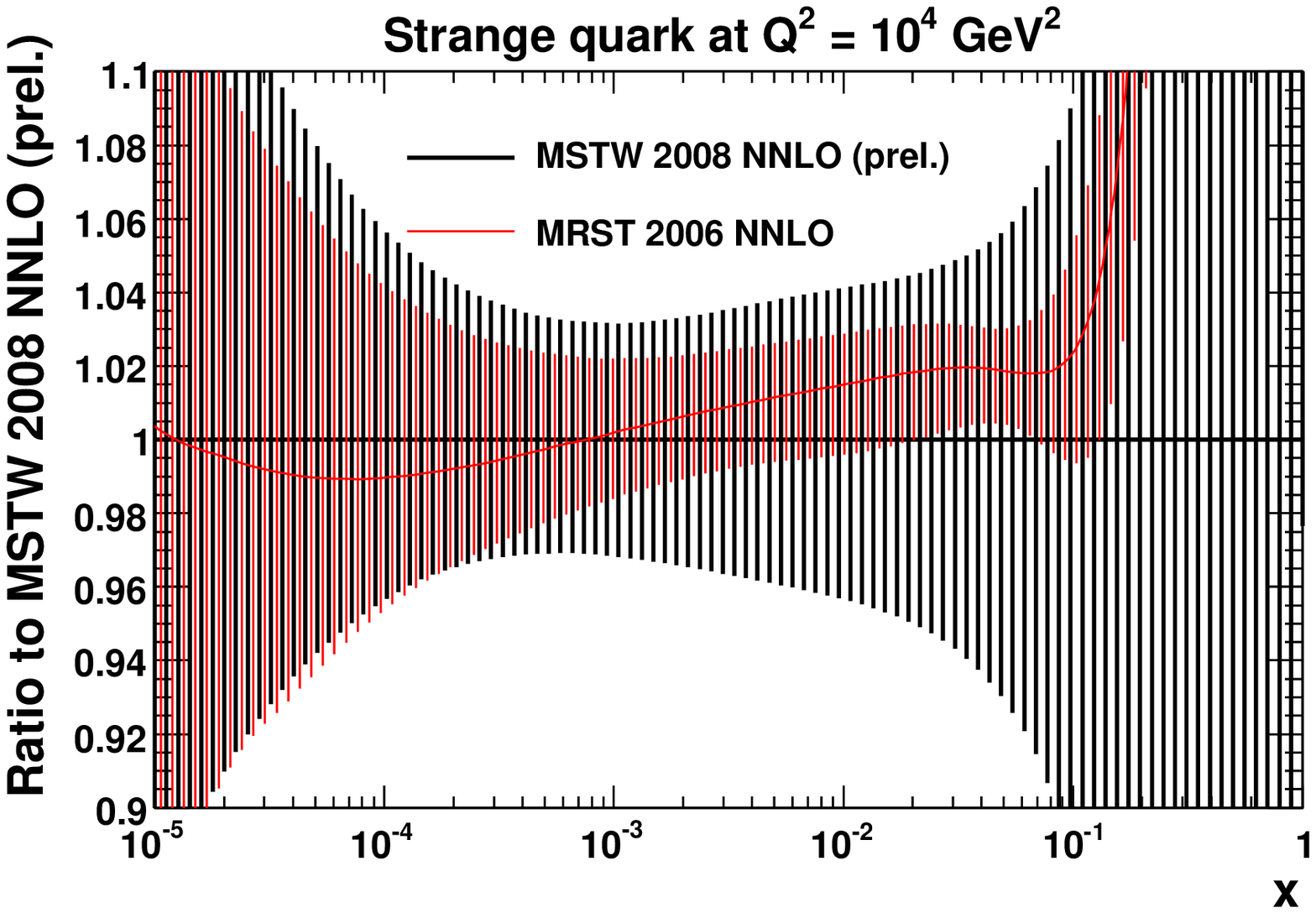}
  \vspace*{-8mm}
  \caption{MSTW 2008 NNLO (prel.) PDFs compared to MRST 2006 NNLO PDFs.}
  \label{fig:comparisonNNLO}
\end{figure}

The treatment of jet data has been improved with respect to the MRST 2001--2006 analyses, where six \emph{pseudogluon} points at $Q^2=2000$ GeV$^2$ inferred from Tevatron Run I inclusive jet data were fitted, rather than the original data points.  The MSTW 2008 analysis uses the \texttt{fastNLO} package to fit the Tevatron Run II and HERA DIS inclusive jet data points, including a complete treatment of the correlated systematic errors.  At NNLO, two-loop threshold corrections are included for Tevatron jet data, and the HERA DIS jet data are excluded.  These developments were reported at DIS 2007~\cite{Thorne:2007bt}, where a fit including CDF Run II jet data was presented.  Recently published D{\O} Run II jet data have now been included in the fit.  The Run II jet data prefer a smaller gluon distribution at high $x$ compared to the previous Run I data, as seen in Fig.~\ref{fig:comparisonNNLO}(a) which shows the ratio of the MRST 2006 NNLO gluon distribution to the new NNLO gluon at $Q^2 = 10^4$ GeV$^2$.

New electroweak data from the Tevatron Run II are now included in the fit.  The lepton charge asymmetry has been measured in $W\to e\nu$ decays in two $E_T^e$ bins by CDF, and in $W\to \mu\nu$ decays in one $p_T^\mu$ bin by D{\O}.  D{\O} will soon have a measurement of the lepton charge asymmetry from $W\to e\nu$ decays in two $p_T^e$ bins, corrected for detector effects, which will be included in the final MSTW 2008 fit.  CDF and D{\O} data on the $Z/\gamma^*$ rapidity distributions are also now included in the fit.  The new electroweak data mainly constrain the down quark distribution since the up quark distribution is already well constrained by structure function data where it appears charge-weighted.  The ratios of the previous MRST 2006 NNLO up and down quark distributions to those from the new fit are shown in Figs.~\ref{fig:comparisonNNLO}(b,c).

The inclusion of NuTeV/CCFR dimuon cross sections allows a more flexible parameterisation of the strange and antistrange distributions, given by \eqref{eq:splus} and \eqref{eq:sminus}, instead of the previous MRST assumption that
$s(x,Q_0^2) = \bar{s}(x,Q_0^2) = \frac{\kappa}{2} \left[\bar{u}(x,Q_0^2) + \bar{d}(x,Q_0^2)\right]$, with $\kappa\approx 0.5$.
The dimuon data prefer a suppression of the strange sea at large $x$ compared to the non-strange sea, as seen in Fig.~\ref{fig:comparisonNNLO}(d).  The new $s$ and $\bar{s}$ distributions have a larger uncertainty than previously since they are no longer tied to $\bar{u}$ and $\bar{d}$ at the input scale.  The best fit gives a non-zero strange sea asymmetry, which is, however, consistent with zero within the 90\% C.L.~limit uncertainty band.

Predictions for the total $W$ and $Z$ cross sections at the Tevatron and LHC are presented in Table \ref{tab:wandz}, together with the ratio of predictions using some other recent PDF sets to those from MSTW 2008.  Note that the change in the total $W$ and $Z$ cross sections going from MRST 2004 to MRST 2006 was due to an improvement in the heavy flavour prescription~\cite{Martin:2007bv}, while the predictions are relatively stable in going from MRST 2006 to MSTW 2008.
\begin{table}
  \centering
  \begin{tabular}{|l|c|c|}
    \hline
    Total $W$ and $Z$ cross sections & $B_{l\nu} \cdot \sigma_W$ (nb) & $B_{l^+l^-}\cdot\sigma_Z$ (nb) \\
    \hline
    MSTW 2008 NLO (prel.) & 20.45 (2.650) & 1.965 (0.2425) \\
    MSTW 2008 NNLO (prel.) & 21.44 (2.739) & 2.043 (0.2512) \\ 
    \hline
  \end{tabular}
  \begin{tabular}{|l|c|c|}
    \multicolumn{3}{c}{}\\\hline
    Ratio to MSTW 2008 (prel.) & $\sigma_W$ & $\sigma_Z$ \\
    \hline
    MRST 2004 NLO~\cite{Martin:2004ir} & 0.974 (0.990) & 0.982 (1.000) \\
    MRST 2004 NNLO~\cite{Martin:2004ir} & 0.936 (0.991) & 0.940 (1.003) \\ \hline
    MRST 2006 NLO (unpublished) & 1.002 (0.995) & 1.009 (1.001) \\
    MRST 2006 NNLO~\cite{Thorne:2007bt} & 0.995 (1.004) & 1.001 (1.010) \\ \hline
    CTEQ6.6 NLO~\cite{Nadolsky:2008zw} & 1.019 (0.978) & 1.022 (0.987) \\
    \hline
  \end{tabular}
  \caption{Predictions for the total $W$ and $Z$ cross sections at the LHC (Tevatron) multiplied by the appropriate branching ratio and calculated in the narrow-width approximation using the PDG 2006 electroweak parameters.} \label{tab:wandz}\vspace{-0.2mm}
\end{table}

We will soon have publically available LO, NLO and NNLO parton distributions, each with 40 additional eigenvector PDF sets, in time for the LHC start-up.

% ****************************************************************************
% BIBLIOGRAPHY AREA
% ****************************************************************************

\begin{footnotesize}

\end{footnotesize}

% ****************************************************************************
% END OF BIBLIOGRAPHY AREA
% ****************************************************************************

\end{document}